\documentstyle[twoside]{article}

\catcode`\@=11
\long\def\@makefntext#1{
\protect\noindent \hbox to 3.2pt {\hskip-.9pt  
$^{{\eightrm\@thefnmark}}$\hfil}#1\hfill}		

\def\@makefnmark{\hbox to 0pt{$^{\@thefnmark}$\hss}}	
	
\def\ps@myheadings{\let\@mkboth\@gobbletwo
\def\@oddhead{\hbox{}
\rightmark\hfil\eightrm\thepage}   
\def\@oddfoot{}\def\@evenhead{\eightrm\thepage\hfil
\leftmark\hbox{}}\def\@evenfoot{}
\def\sectionmark##1{}\def\subsectionmark##1{}}

\oddsidemargin=\evensidemargin
\newcounter{sectionc}\newcounter{subsectionc}\newcounter{subsubsectionc}
\renewcommand{\section}[1] {\vspace{12pt}\addtocounter{sectionc}{1} 
\setcounter{subsectionc}{0}\setcounter{subsubsectionc}{0}\noindent 
	{\tenbf\thesectionc. #1}\par\vspace{5pt}}
\renewcommand{\subsection}[1] {\vspace{12pt}\addtocounter{subsectionc}{1} 
	\setcounter{subsubsectionc}{0}\noindent 
	{\bf\thesectionc.\thesubsectionc. {\kern1pt \bfit #1}}\par\vspace{5pt}}
\renewcommand{\subsubsection}[1] {\vspace{12pt}\addtocounter{subsubsectionc}{1}
	\noindent{\tenrm\thesectionc.\thesubsectionc.\thesubsubsectionc.
	{\kern1pt \tenit #1}}\par\vspace{5pt}}

\newcounter{appendixc}
\newcounter{subappendixc}[appendixc]
\newcounter{subsubappendixc}[subappendixc]
\renewcommand{\thesubappendixc}{\Alph{appendixc}.\arabic{subappendixc}}
\renewcommand{\thesubsubappendixc}
	{\Alph{appendixc}.\arabic{subappendixc}.\arabic{subsubappendixc}}

\renewcommand{\appendix}[1] {\vspace{12pt}
        \refstepcounter{appendixc}
        \setcounter{figure}{0}
        \setcounter{table}{0}
        \setcounter{lemma}{0}
        \setcounter{theorem}{0}
        \setcounter{corollary}{0}
        \setcounter{definition}{0}
        \setcounter{equation}{0}
        \renewcommand{\thefigure}{\Alph{appendixc}.\arabic{figure}}
        \renewcommand{\thetable}{\Alph{appendixc}.\arabic{table}}
        \renewcommand{\theappendixc}{\Alph{appendixc}}
        \renewcommand{\thelemma}{\Alph{appendixc}.\arabic{lemma}}
        \renewcommand{\thetheorem}{\Alph{appendixc}.\arabic{theorem}}
        \renewcommand{\thedefinition}{\Alph{appendixc}.\arabic{definition}}
        \renewcommand{\thecorollary}{\Alph{appendixc}.\arabic{corollary}}
        \renewcommand{\theequation}{\Alph{appendixc}.\arabic{equation}}
        \noindent{\tenbf Appendix \theappendixc #1}\par\vspace{5pt}}
\newcommand{\subappendix}[1] {\vspace{12pt}
        \refstepcounter{subappendixc}
        \noindent{\bf Appendix \thesubappendixc. {\kern1pt \bfit #1}}
	\par\vspace{5pt}}
\newcommand{\subsubappendix}[1] {\vspace{12pt}
        \refstepcounter{subsubappendixc}
        \noindent{\rm Appendix \thesubsubappendixc. {\kern1pt \tenit #1}}
	\par\vspace{5pt}}

\newcommand{\textlineskip}{\baselineskip=13pt}
\newcommand{\smalllineskip}{\baselineskip=10pt}

\def\eightcirc{
\begin{picture}(0,0)
\put(4.4,1.8){\circle{6.5}}
\end{picture}}
\def\eightcopyright{\eightcirc\kern2.7pt\hbox{\eightrm c}} 

\newcommand{\copyrightheading}[1]
	{\vspace*{-2.5cm}\smalllineskip{\flushleft
	{\footnotesize International Journal of Modern Physics A, #1}\\
	{\footnotesize $\eightcopyright$\, World Scientific Publishing
	 Company}\\
	 }}

\def\abstracts#1#2#3{{
	\centering{\begin{minipage}{4.5in}\baselineskip=10pt\footnotesize
	\parindent=0pt #1\par 
	\parindent=15pt #2\par
	\parindent=15pt #3
	\end{minipage}}\par}} 

\newcommand{\bibit}{\nineit}

\renewenvironment{thebibliography}[1]
	{\frenchspacing
	 \ninerm\baselineskip=11pt
	 \begin{list}{\arabic{enumi}.}
	{\usecounter{enumi}\setlength{\parsep}{0pt}
	 \setlength{\leftmargin 12.7pt}{\rightmargin 0pt} 
	 \setlength{\itemsep}{0pt} \settowidth
	{\labelwidth}{#1.}\sloppy}}{\end{list}}

\newcounter{itemlistc}
\newcounter{romanlistc}
\newcounter{alphlistc}
\newcounter{arabiclistc}

\newcommand{\fcaption}[1]{
        \refstepcounter{figure}
        \setbox\@tempboxa = \hbox{\footnotesize Fig.~\thefigure. #1}
        \ifdim \wd\@tempboxa > 5in
           {\begin{center}
        \parbox{5in}{\footnotesize\smalllineskip Fig.~\thefigure. #1}
            \end{center}}
        \else
             {\begin{center}
             {\footnotesize Fig.~\thefigure. #1}
              \end{center}}
        \fi}

\newcommand{\tcaption}[1]{
        \refstepcounter{table}
        \setbox\@tempboxa = \hbox{\footnotesize Table~\thetable. #1}
        \ifdim \wd\@tempboxa > 5in
           {\begin{center}
        \parbox{5in}{\footnotesize\smalllineskip Table~\thetable. #1}
            \end{center}}
        \else
             {\begin{center}
             {\footnotesize Table~\thetable. #1}
              \end{center}}
        \fi}

\def\@citex[#1]#2{\if@filesw\immediate\write\@auxout
	{\string\citation{#2}}\fi
\def\@citea{}\@cite{\@for\@citeb:=#2\do
	{\@citea\def\@citea{,}\@ifundefined
	{b@\@citeb}{{\bf ?}\@warning
	{Citation `\@citeb' on page \thepage \space undefined}}
	{\csname b@\@citeb\endcsname}}}{#1}}

\newif\if@cghi
\def\cite{\@cghitrue\@ifnextchar [{\@tempswatrue
	\@citex}{\@tempswafalse\@citex[]}}
\def\citelow{\@cghifalse\@ifnextchar [{\@tempswatrue
	\@citex}{\@tempswafalse\@citex[]}}
\def\@cite#1#2{{$\null^{#1}$\if@tempswa\typeout
	{IJCGA warning: optional citation argument 
	ignored: `#2'} \fi}}

\def\pmb#1{\setbox0=\hbox{#1}
	\kern-.025em\copy0\kern-\wd0
	\kern.05em\copy0\kern-\wd0
	\kern-.025em\raise.0433em\box0}

\def\fnt#1#2{\footnotetext{\kern-.3em
	{$^{\mbox{\scriptsize #1}}$}{#2}}}

\def\fpage#1{\begingroup
\voffset=.3in
\thispagestyle{empty}\begin{table}[b]\centerline{\footnotesize #1}
	\end{table}\endgroup}

\def\runninghead#1#2{\pagestyle{myheadings}
\markboth{{\protect\footnotesize\it{\quad #1}}\hfill}
{\hfill{\protect\footnotesize\it{#2\quad}}}}
\font\tenrm=cmr10
\font\tenit=cmti10 
\font\tenbf=cmbx10
\font\bfit=cmbxti10 at 10pt
\font\ninerm=cmr9
\font\nineit=cmti9

\font\eightrm=cmr8

\def\qed{\hbox{${\vcenter{\vbox{			
   \hrule height 0.4pt\hbox{\vrule width 0.4pt height 6pt
   \kern5pt\vrule width 0.4pt}\hrule height 0.4pt}}}$}}

\def\d{\partial}
\def\<{\langle}
\def\>{\rangle}
\def\x{\mbox{\boldmath$x$}}
\def\y{\mbox{\boldmath$y$}}

\begin{document}

\runninghead{Hydrodynamics of relativisic systems with broken
continuous symmetries} {Hydrodynamics of relativisic systems with
broken continuous symmetries}

\normalsize\textlineskip
\thispagestyle{empty}
\setcounter{page}{1}

\copyrightheading{}			

\vspace*{0.88truein}

\fpage{1}
\centerline{\bf HYDRODYNAMICS OF RELATIVISTIC SYSTEMS}
\vspace*{0.035truein} \centerline{\bf WITH BROKEN CONTINUOUS
SYMMETRIES
}
\vspace*{0.37truein}
\centerline{\footnotesize D.T.~SON
}
\vspace*{0.015truein}
\centerline{\footnotesize\it Physics Department, Columbia University, 528 West 120th St.}
\baselineskip=10pt 
\centerline{\footnotesize\it New York, New York
10027, USA
}
\vspace*{0.225truein}

\vspace*{0.21truein} 
\abstracts{We show how hydrodynamics of relativistic system with
broken continuous symmetry can be constructed using the Poisson
bracket technique.  We illustrate the method on the example of
relativistic superfluids.}{}{}{}

\textlineskip			
\vspace*{12pt}			

\noindent
The study of hydrodynamics of relativistic systems is important for
the physics of heavy-ion collisions.  Indeed, hydrodynamic models are
the conceptually simplest models of heavy-ion
reactions.\cite{Landau,Bjorken}
 
By definition, hydrodynamics is the effective theory describing the
real-time dynamics of a given system at length scales larger than the
mean free path and time scales larger than the mean free time.
Implicitly in this definition, one assumes a nonzero temperature for
the mean free path to be finite.  At large length and time scales,
only a small number of degrees of freedom survive to become
hydrodynamic modes.  If the system is far away from all second order
phase transitions, the latter modes are divided into two categories:
the conserved densities (i.e.\ densities of conserved charged), and
the phases of order parameters.  In systems with no broken continuous
symmetry, all hydrodynamics modes belong to the first type, and the
hydrodynamic equations are the conservation laws.  But if there are
broken continuous symmetries in the theory, then they give rise to new
hydrodynamic modes.  Our task is to find out the hydrodynamic
equations that describe the coupled dynamics of Goldstone and fluid
dynamical modes.  This can be done using the Poisson bracket method
described below.
 
For definiteness we will consider the simplest theory with a broken
symmetry, the complex $\phi^4$ field theory:
\[
  L = (\d_\mu\phi^*)(\d_\mu\phi) - \lambda(|\phi|^2 -v^2)^2
\]
This theory is the relativistic generalization of superfluids.  The
set of hydrodynamic modes in this theory include five conserved
densities: the energy density $T^{00}$, three components of the
momentum density $T^{0i}$, and the density of the U(1) charge
$n=-{i\over2}(\phi^*\d_0\phi-\d_0\phi^*\phi)$, and one non-conserved
phase $\varphi$, $\< \phi \> = |\< \phi \>| e^{i\varphi}$.  If one
ignore dissipative processes, then instead of $T^{00}$ one can use the
entropy density $s$.

At the operational level, the Poisson bracket method works as
follows.  We first write down the classical Poisson bracket between
the hydrodynamic variables
\begin{eqnarray}
  {}[ T^{0i}(\x),\, T^{0k}(\y) ] & = & 
     \biggl[ T^{0k}(\x){\d\over\d x^i} - T^{0i}(\y){\d\over\d y^k}
     \biggr] \delta(\x-\y) \label{TT} \\
  {}[ T^{0i}(\x),\, n(\y) ] & = & n(\x)\d_i\delta(\x-\y) \label{Tn}\\
  {}[ T^{0i}(\x), s(\y) ] & = & s(\x) \d_i\delta(\x-\y) \label{Ts}\\
  {}[ T^{0i}(\x), \varphi(\y) ] & = & \d_i\varphi \delta(\x-y) \label{Tphi}\\
  {}[ n(\x), \varphi(\y) ] & = & - \delta(\x-\y) \label{nphi}
\end{eqnarray}
Eqs.\ (\ref{TT},\ref{Tn}) can be derived by direct calculation of
quantum commutators, using the canonical commutation relation.  Eqs.\
(\ref{Ts},\ref{Tphi},\ref{nphi}) are postulated from physical
considerations.  Eqs.\ (\ref{Ts},\ref{Tphi}) is consistent with the
fact that $\int\!d\x\, T^{0i}(\x)$ is the total momentum and generates
a coordinate translation.  The difference in the form of Eqs.\
(\ref{Ts}) and (\ref{Tphi}) is due to the different dimensionalities
of $s$ and $\phi$, which transform differently under dilatation
$\int\!d\x\,x^iT^{0i}(\x)$.  Finally, Eq.\ (\ref{nphi}) tells us that
the charge and the phase of the condensate are conjugate variables.

The next step is to write down the most general Hamiltonian, which is
also the total energy,
\begin{equation}
  H = \int\! d\x\, T^{00}(s, n, T^{0i}, \d_i\varphi)
  \label{H}
\end{equation}
Note that $T^{00}$ may depends only on the derivatives of $\varphi$
but not on $\varphi$ itself due to the invariance under the U(1)
rotation.  The form of the function $T^{00}$ is undefined at this
moment and can be completely found only by calculations at the
microscopic level.  However some constraint on the form of $T^{00}$
follows from relativistic invariance, see below.  Knowing the
Hamiltonian (\ref{H}) and the Poisson brackets (\ref{TT}-\ref{nphi}),
the classical dynamics of the hydrodynamic modes are fixed completely.
The equation of motion for any variable $A$ has the form
\begin{equation}
  \dot A = [H, A]
\end{equation}
In particular, one can find the equation of motion for the energy
density $T^{00}$, $\dot T^{00} = [H, T^{00}]$.  From relativistic
invariance one expects that $\dot T^{00} = -\d_i T^{0i}$: the energy
flux is equal to the momentum density.  This imposes a severe
constraint on the possible forms of the function $T^{00}(s, n, T^{0i},
\d_i\varphi)$, and makes possible a relativistically covariant
formulation of hydrodynamics.  Omitting computational details, the
final formulation is as follows.

First, one should find, from microscopic physics, the equation of
state, which define the pressure as a function of three variables,
\begin{equation}
  p = p (T, \mu, {1\over2}(\d_\mu\varphi)^2 )
  \label{eqstate}
\end{equation}
where the first two variables are nothing but the temperature and the
chemical potential of the conserved charge, and the third variable is
specific for the superfluid phase and defines the degree of variation
of the U(1) condensate phase over space-time.  Knowing the equation of
state (\ref{eqstate}), one defines the thermodynamic variables
conjugate to $T$, $\mu$ and ${1\over2}(\d_\mu\varphi)$:
\begin{equation}
  dp = sdT + nd\mu + V^2 d({1\over2}(\d_\mu\varphi)^2)
\end{equation}
One also defines $\rho$ as the Legendre transform of $-p$ with respect
to $T$ and $\mu$, $\rho = sT + n\mu-p$.  The physical meaning of $s$,
$n$ and $V$ is clear from the hydrodynamic equations which have the
form
\begin{eqnarray}
  & & \d_\mu T^{\mu\nu} = 0, \qquad 
    T^{\mu\nu} = (\rho+p)u^\mu u^\nu - pg^{\mu\nu} + 
    V^2\d^\mu\varphi\d^\nu\varphi \label{energy_cons}\\
  & & \d_\mu (nu^\mu - V^2 \d^\mu\varphi) = 0 \label{n_cons}\\
  & & \d_\mu (su^\mu) = 0 \label{s_cons} \\
  & & u^\mu \d_\mu\varphi + \mu =0 \label{Josephson}
\end{eqnarray}
Eq.\ (\ref{energy_cons}) is energy-momentum conservation, where the
energy-momentum tensor $T^{\mu\nu}$ consists of two parts: a ``fluid''
part that is due to the particles outside the condensate, and a
``field'' part which is due to the coherent motion of the condensate.
Similarly, Eq.\ (\ref{n_cons}) is conservation of the U(1) charge,
where the conserved current is also a sum of a normal current and a
superfluid current.  The velocity $u^\mu$ is clearly the velocity of
the normal component of the fluid.  Eq.\ (\ref{s_cons}) is the
conservation law for entropy: only the normal component contributes to
the entropy flow, in accordance with physical expectations.  Finally
Eq.\ (\ref{Josephson}) is not a conservation law but a Josephson-type
equation that relate the time derivative of the condensate phase in
the frame where no normal flow occurs with the chemical potential.

It can be shown that Eqs.\ (\ref{energy_cons}-\ref{Josephson}) are
equivalent (although only after a nontrivial mapping) to the set of
equations previously suggested by Carter, Khalatnikov and Lebedev for
relativistic superfluids.\cite{KL,CK} The advantage of the form
(\ref{energy_cons}-\ref{Josephson}) is that the physical meaning and
the field-theoretical origin of all quantities and equations are made
quite clear.

One can also generalize this technique to the case of nuclear matter.
Near the chiral limit, nuclear matter possesses a broken continuous
symmetry, which is $SU(N_f)\times SU(N_f)\to SU(N_f)$.  The
hydrodynamics in this case described the evolution of $5+2(N_f^2-1)$
conserved charges (energy and momentum, entropy, and left and
right-handed isospin charges) and $N^2-1$ condensate phases (the
Goldstone bosons).  For details, see Ref.\ \cite{Son}.

\end{document}